%% file: main.tex
\documentclass[conference]{IEEEtran}
\IEEEoverridecommandlockouts
\usepackage{cite}
\usepackage{amsmath,amssymb,amsfonts, amsthm}
\usepackage{graphicx}
\usepackage{textcomp}
\usepackage{mathtools}
\usepackage{color,soul}
\usepackage{algorithm}
\usepackage{algpseudocode}
\usepackage[most]{tcolorbox}

\usepackage{booktabs}
\usepackage{multirow}

\newtheorem{theorem}{Theorem}[section]

\newtheorem{problem}[theorem]{Problem}

\newtheorem{definition}{Definition}[section]

\definecolor{rqbg}{RGB}{232,238,255} 

\newtcolorbox{rqsummarybox}{
  colback=rqbg,
  colframe=rqbg,   
  boxrule=0pt,
  arc=0pt,         
  boxsep=0pt,
  left=5pt,right=5pt,top=4pt,bottom=4pt,
  width=\linewidth,
  before skip=6pt,
  after skip=6pt,
}

\def\BibTeX{{\rm B\kern-.05em{\sc i\kern-.025em b}\kern-.08em
    T\kern-.1667em\lower.7ex\hbox{E}\kern-.125emX}}
\begin{document}

\title{No Tile Left Behind: Multiprogramming for Surface-Code Architectures
}

\author{
\IEEEauthorblockN{
Archisman Ghosh\IEEEauthorrefmark{1},
Avimita Chatterjee\IEEEauthorrefmark{1},
Swaroop Ghosh\IEEEauthorrefmark{3}
}

\IEEEauthorblockA{\IEEEauthorrefmark{1}\textit{Department of Computer Science and Engineering}, The Pennsylvania State University}
\IEEEauthorblockA{\IEEEauthorrefmark{3}\textit{School of Electrical Engineering and Computer Science}, The Pennsylvania State University}
}

\maketitle

\begin{abstract}
Fault-tolerant quantum computing (FTQC) is emerging as the architectural regime in which practical large-scale quantum workloads will execute. In this setting, however, multiprogramming is no longer a matter of partitioning a flat pool of qubits. Quantum error correction exposes a structured floorplan of data tiles, ancilla tiles, and magic-state service resources, so concurrent execution must account for compact placement, connectivity, routing headroom, and shared support infrastructure. This makes FTQC multiprogramming fundamentally harder than its NISQ counterpart: admission decisions can fragment the remaining floorplan, conservative reservations can waste ancilla, and dynamic contention across data, ancilla, and magic-state resources can degrade both throughput and quality of service. In this work, we develop a formal framework for FTQC multiprogramming that captures these structural constraints and their runtime implications. We formulate the baseline static allocation problem, extend it to limited-resource and online settings through hierarchy-aware scheduling policies, and further generalize it to cultivation-enabled architectures with dynamic magic-state generation. Through simulation on synthetic Clifford+T workloads, the proposed scheduler achieves a normalized system speedup of $3.1\times$, improving over prior FTQC multiprogramming baselines by $\sim29\%$ while maintaining low mean slowdown. 
\end{abstract}


\input{1_introduction.tex}
\input{2_background.tex}
\input{3_problem.tex}
\input{4_cultivation.tex}
\input{5_evaluation.tex}

\input{6_Discussion.tex}

\input{7_conclusion.tex}

\section{Acknowledgements}
\noindent The work is supported in parts by the National Science Foundation (NSF) (CNS-1722557, CCF-1718474) and gifts from Intel. 

\bibliography{refs}
\bibliographystyle{ieeetr}
\end{document}

%% file: 1_introduction.tex
\section{Introduction}

\begin{figure*}
    \centering
    \includegraphics[width=1\linewidth]{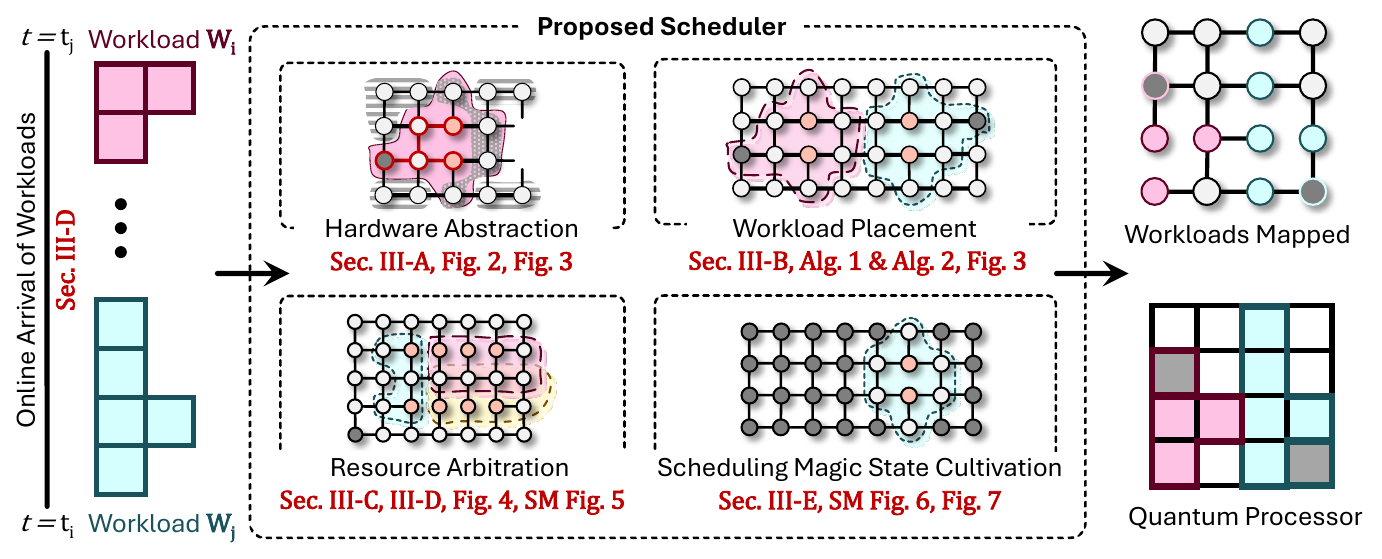}
    \caption{\textbf{Overview of the Proposed Multiprogramming Scheduler.} Diagrammatic summary of our end-to-end framework for multiprogramming in surface-code FTQC. Online workloads are processed by the proposed scheduler, which combines hardware abstraction, workload placement, resource arbitration, and magic-state cultivation scheduling. We present our proposed framework in Section \ref{sec:multi}. The output is a mapped workload graph, equivalent to the occupied quantum-processor layout. The figure highlights the paper’s key contributions and the sections where each component is formalized; SM denotes State Machine.}
    \label{fig:Contribution}
\end{figure*}

Quantum computing is transitioning from the noisy intermediate-scale quantum (NISQ) regime toward a fault-tolerant regime with an increased number of better-quality physical qubits with Quantum Error Correction (QEC) to protect logical qubits \cite{Bravyi_2024, gidney2024magicstatecultivationgrowing}. Fault tolerant quantum computers (FTQC) can provide the quantum advantage over classically intractable problems. 
With increasing research in FTQC, public roadmaps and vendor claims suggest access to large-scale FTQC hardware with specific qubit technologies to be available by the end of this decade \cite{atom_ac1000_2026,  atom_preview_2026, ibm_ftqc_2025}. As fault-tolerant quantum hardware scales toward hundreds of logical qubits, cloud providers will face scheduling problems: How to share a structured surface-code floorplan? How to reason the admission of online workloads? How to schedule magic state resources?  
This makes concurrent execution of jobs (\emph{multiprogramming}) on the FTQC hardware an important avenue for research, as it maximally utilizes available physical qubits. While existing research in NISQ-era quantum computers has explored the scope of multiprogramming, FTQC presents a different set of challenges.

\noindent\textbf{Difference in abstraction: }On one hand, hardware abstraction in NISQ computers is merely the topology of the physical qubits. The workloads are transpiled into the native gate set of the hardware, and a scheduler has to deal with device-level constraints like connectivity, calibration, and crosstalk errors. The workloads also have to maintain good fidelity while being executed concurrently. On the other hand, in FTQC architectures, with the presence of a QEC layer between the hardware and the workload, the computation shifts to the logical qubits, which are now protected by a number of lower-error physical qubits through a QEC code. The scheduler in this case has to account for the abstraction changes from a noisy qubit graph to a structured logical floorplan with an underlying QEC code. For surface-code architectures, the floorplan is expressed as a grid of logical data patches, ancilla tiles for measurement, and magic state factories for implementing $T$ gates. Therefore, placing multiple workloads in such a floorplan becomes a multi-constrained spatio-temporal mapping problem where the scheduler needs to take the shared ancilla and magic state resources into account.

\noindent\textbf{Difference in optimization goals: }
With the shift in abstraction, the optimization objective of a multiprogramming scheduler also changes. In NISQ systems, the primary goal is to execute the workload successfully by tackling noise in the system through various techniques like reducing the circuit depth, \textsc{SWAP} gates, or less interference between concurrently executing jobs. However, in FTQCs, the qubits have less noise due to the QEC layer, and hence, performance is governed less by device noise and more by efficient use of logical space-time resources. The key optimization goals in FTQC systems thus become logical area utilization, fragmentation of the floorplan, and effective sharing of resources (ancilla tiles and magic state factories). A circuit-level schedule obtained from NISQ multiprogramming principles may be suboptimal if it fragments the floorplan, starves the workload of magic state resources, and/or reduces throughput.

\emph{In this paper, we develop a formal abstraction for fault-tolerant surface-code architectures and demonstrate efficient resource sharing in resource-constrained settings with online workload arrivals.} Our contributions are as follows (Fig. \ref{fig:Contribution}):
\begin{enumerate}
    \item We show that FTQC multiprogramming is a multi-constrained spatio-temporal placement problem and provide a formal graph-theoretic definition that captures data, ancilla, and magic-state constraints jointly (Section \ref{subsec:hw-abstraction-notation}).
    
    \item We design a heuristic solution to minimize resource wastage and develop hierarchy-aware policies for limited-resource scenarios in which data tiles, magic-state resources, or ancilla capacity become the dominant bottleneck (Section \ref{subsec:resource-sufficient-static-allocation}, \ref{subsec:limitedresource}).

    \item Finally, we explore the scope of accommodating magic state cultivation-enabled architectures, in which fixed magic state ports are replaced by dynamic magic state generation on ancilla tiles (Section \ref{sec:cultivation-enabled}). 

    \item We numerically evaluate our proposed scheduling policies on a diverse set of synthetic benchmarks and obtain an average normalized throughput of $3.1\times$ relative to sequential standalone execution, and $\sim29\%$ improvement over the only prior FTQC multiprogramming baseline (Section \ref{sec:eval}).
    
\end{enumerate}
Section \ref{sec:related} describes the related work and Section \ref{sec:conc} concludes the paper.

%% file: 2_background.tex
\section{Surface Code Architectures}

Surface codes are widely studied for fault-tolerant quantum computation due to high error thresholds and local 2-D connectivity \cite{Fowler_2012, Horsman_2012}. We use a planar surface-code architecture \cite{Litinski_2018,Litinski_2019,ghosh2025designautomationquantumerror, lsqca, Beverland_2022, ghosh2026designingworkloadawaresurfacecode}, where each logical qubit is a 2-D nearest-neighbor patch. A distance-$d$ patch uses $O(d^2)$ physical qubits and is protected by repeated stabilizer measurements. Its two $X$- and two $Z$-type boundaries define logical operators $\bar{X}$ and $\bar{Z}$ via Pauli strings connecting like boundaries. The code distance is the minimum length of a non-trivial logical operator, with fault tolerance maintained by $d$ syndrome-extraction rounds during storage and reconfiguration. Thus, computation is organized around logical patches and their boundaries rather than individual physical qubits.
Logical interactions are commonly implemented via lattice surgery, which preserves 2-D locality \cite{Chatterjee_2025,dascot,Horsman_2012}. Rather than using long-range couplings or extensive \textsc{SWAP} networks, lattice surgery performs computation through local boundary deformations between adjacent patches. Its merge and split operations enable joint Pauli measurements between neighboring logical patches, forming the basis for patch-based logical operations \cite{Litinski_2018}. Thus, patch placement determines which interactions are efficient and which require movement, reshaping, or temporary workspace.
At the architectural level, we model the surface-code floorplan as a grid of logical tiles,
\(
\mathcal{T}=\{T_{i,j}\},
\)
where each tile represents a fixed surface-code footprint at distance $d$. Logical qubits are represented by surface-code patches placed on these tiles. A one-qubit patch may occupy a single tile or a small connected group of tiles depending on the supported operation, while larger multi-tile patches may be used to expose particular boundary configurations or enable more efficient lattice-surgery measurements. This tile-and-patch abstraction provides a logical-level description of the physical floorplan while retaining the geometric locality constraints imposed by the surface code.

We distinguish three primary architectural resources. First, \emph{data patches} are long-lived logical patches that store the logical qubits of the computation. Their placement determines which joint Pauli measurements can be directly supported by local lattice surgery and which require patch movement or reconfiguration. Second, \emph{ancilla} or \emph{compute} regions provide temporary workspace for lattice-surgery operations, syndrome extraction, and intermediate measurement structures. Third, \emph{magic-state resources}, implemented using dedicated distillation blocks or factories, supply high-fidelity magic states required to realize non-Clifford gates~\cite{Litinski_2019_msd,Bravyi_2012}. Together, these resources define the spatial and temporal constraints under which a fault-tolerant workload must execute.

Prior work on surface-code architecture design has largely focused on efficiently executing a single fault-tolerant workload \cite{silva, dascot}. These studies have investigated patch layouts, lattice-surgery schedules, routing strategies, and magic-state provisioning for reducing the space-time cost of an individual computation \cite{ftqc_sync, swiper, ravi2022betterworstcasedecodingquantum}. However, future fault-tolerant quantum systems are unlikely to operate exclusively in a single-program mode. 
As hardware scales, multiple independent workloads will share the same surface-code fabric, shifting the challenge from optimizing a single computation to managing shared data, ancilla, and magic-state resources across concurrent workloads. This motivates a multiprogramming framework that reduces the resource overhead, and improves execution efficiency.

%% file: 3_problem.tex
\section{The Multiprogramming Approach}\label{sec:multi}

\subsection{Hardware Abstraction and Notation}
\label{subsec:hw-abstraction-notation}

\begin{figure}
    \centering
    \includegraphics[width=1\linewidth]{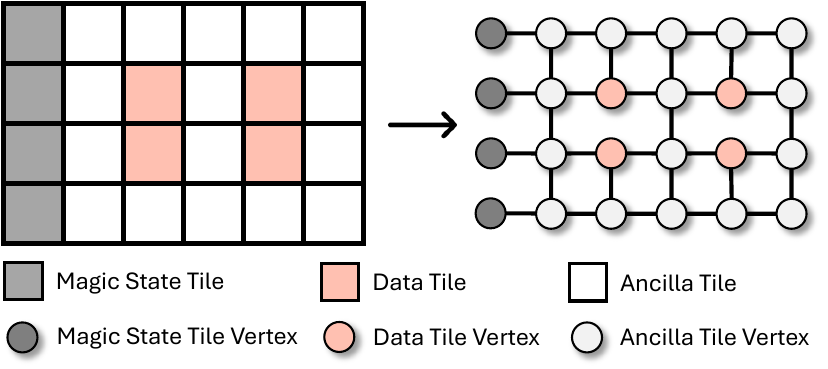}
    \caption{\textbf{Hardware Abstraction.} An example layout comprising four data tiles and four magic state ports. The FTQC abstraction for surface code architecture can be represented as a 2D nearest-neighbor graph, as shown.}
    \label{fig:3a_ftqc_notation}
\end{figure}

We abstract the fault-tolerant surface-code architecture as an unweighted undirected graph
\(
G=(V,E)
\)
(Fig. \ref{fig:3a_ftqc_notation}) where the vertices represent the tiles (ancilla, data patches, and magic state ports), and edges represent nearest-neighbor adjacency available for lattice-surgery interaction, patch movement, or routing through the layout. For any two vertices \(u,v\in V\), let
\(
d_G(u,v)
\)
denote the shortest-path distance between \(u\) and \(v\) in \(G\). For any subset \(U\subseteq V\), we write \(G[U]\) for the subgraph of \(G\) induced by \(U\).
The vertex set is partitioned into three pairwise disjoint classes
\(
V = V_D \sqcup V_A \sqcup V_M,
\)
where \(V_D\) is the set of data tiles, \(V_A\) is the set of ancilla tiles, and \(V_M\) is the set of magic-state ports as fixed locations through which high-fidelity magic states are injected for computation.
This port-based abstraction is the default architectural model used throughout the present section.
We consider a finite set of workloads
\(
\mathcal{W}=\{W_1,W_2,\dots,W_K\}.
\)
A spatial allocation for each workload \(W_i\) is represented by a connected subgraph
\(
H_i=(V_i,E_i),
\)
where
\(
V_i = D_i \cup A_i \cup \{m_i\}.
\)
Here, \(D_i \subseteq V_D\) is the set of data vertices assigned to workload \(W_i\), \(A_i \subseteq V_A\) is the set of ancilla vertices committed to the workload's resident core allocation, and \(m_i\in V_M\) is the magic state port vertex associated with the workload.



We distinguish the available ancilla between \emph{core ancilla}, \emph{primary scratchpad ancilla}, and \emph{secondary scratchpad ancilla}. For workload \(W_i\), the core ancilla is exactly the set \(A_i\). Once the resident core \(H_i\) has been fixed, its primary scratchpad is defined as 
\(
P_i
=
\{
a \in V_A \setminus A_i : d_G(a,V_i)=1
\},
d_G(a,V_i):=\min_{v\in V_i} d_G(a,v).
\)
Therefore, \(P_i\) consists of ancilla tiles adjacent to the resident allocation.
The residual ancilla remaining in the architecture outside the workload core forms the secondary scratchpad,
\(
S
=
V_A \setminus \bigcup_{i=1}^K A_i.
\)
Under this abstraction, the multiprogramming task is to assign data, ancilla, and magic-state resources to multiple workloads on the shared floorplan. We next formalize this task as a static optimization problem.

\subsection{Resource-Sufficient Static Allocation}
\label{subsec:resource-sufficient-static-allocation}

We first consider the \emph{resource-sufficient static regime}, which serves as the baseline formulation for the rest of the paper. In this regime, the aggregate data and ancilla vertices are assumed sufficient for the set of workloads under consideration, with a magic-state-port fixed to each workload. Therefore, the central question is to construct a feasible family of simultaneous resident allocations that preserves locality and minimizes unnecessary commitment of ancilla.


\begin{definition}[]
\label{def:growth-admissible-cluster}
For a fixed workload \(W_i\) with data demand \(q_i\) and fixed magic-state port \(m_i\in V_M\), a set
\(
D_i=\{d_i^{(1)},d_i^{(2)},\dots,d_i^{(q_i)}\}\subseteq V_D
\)
is called a \emph{compact data cluster} for \(W_i\) if its vertices admit an ordering such that
\(
d_i^{(1)} \in \arg\min_{v\in V_D} d_G(m_i,v),
\)
and for every \(t=2,\dots,q_i\),
\[
d_i^{(t)}
\in
\arg\min_{v\in V_D\setminus \{d_i^{(1)},\dots,d_i^{(t-1)}\}}
d_G\!(v,\{d_i^{(1)},\dots,d_i^{(t-1)}\}),
\]
where for any set \(U\subseteq V\),
\(
d_G(v,U):=\min_{u\in U} d_G(v,u).
\)
Therefore, the cluster can be generated by seeding at the data vertex nearest to the designated magic-state port and then repeatedly adding the currently unoccupied data vertex closest to the existing partial cluster.
\end{definition}




\begin{problem}\label{prob:main}
Under the hardware abstraction and notation introduced above, the goal is to find a family of resident workload subgraphs
\(
\mathcal{H}=\{H_1,...,H_K\},
\)
such that, for every workload \(W_i\),
\begin{enumerate}
    \item \(D_i\subseteq V_D\), \(A_i\subseteq V_A\), and \(|D_i|=q_i\);
    \item \(D_i\) is a compact data cluster in the sense of Definition~\ref{def:growth-admissible-cluster};
    \item \(H_i\) is connected;
    \item Magic state ports are pairwise distinct, i.e., \(m_i \neq m_j, \forall, i \neq j\);
    \item \(V_i\cap V_j=\varnothing\) for all \(i\neq j\).
\end{enumerate}
The objective is to minimize the total number of ancilla vertices committed to workloads:
\[
\min_{\mathcal{H}} \sum_{i=1}^K |A_i|.
\]
\end{problem}

We propose a two-step solution to Problem \ref{prob:main}: (a) compact partitioning of data vertices, and (b) constructing a core subgraph rooted at the magic-state port.
\begin{algorithm}[t]
\caption{Greedy Compact Partitioning of Data Vertices}
\label{alg:greedy-compact-partition}
\fontsize{8.5pt}{9.5pt}\selectfont
\begin{algorithmic}[1]
\Require Floorplan graph \(G=(V,E)\) with data vertices \(V_D\), demands \((q_1,\dots,q_K)\), fixed ports \((m_1,\dots,m_K)\)
\Ensure Pairwise disjoint data clusters \(D_1,\dots,D_K \subseteq V_D\) with \(|D_i|=q_i\) for all \(i\), if successful
\State \(U_D \gets V_D\)
\For{\(i=1\) to \(K\)}
    \State \(D_i \gets \varnothing\)
    \If{\(q_i > 0\)}
        \State choose \(d_i^{(1)} \in \arg\min_{v\in U_D} d_G(m_i,v)\)
        \State \(D_i \gets \{d_i^{(1)}\}\), \quad \(U_D \gets U_D \setminus \{d_i^{(1)}\}\)
    \EndIf
\EndFor
\While{there exists \(i\) with \(|D_i|<q_i\)}
    \For{\(i=1\) to \(K\)}
        \If{\(|D_i|<q_i\)}
            \If{\(U_D=\varnothing\)}
                \State \Return Failure
            \EndIf
            \State choose \(d_i^\star \in \arg\min_{v\in U_D} \min_{u\in D_i} d_G(v,u)\)
            \State \(D_i \gets D_i \cup \{d_i^\star\}\), \quad \(U_D \gets U_D \setminus \{d_i^\star\}\)
        \EndIf
    \EndFor
\EndWhile
\State \Return \((D_1,\dots,D_K)\)
\end{algorithmic}
\end{algorithm}

\noindent\textbf{Greedy compact partitioning of data vertices:}
We first assign pairwise disjoint data clusters to the workloads. For workload \(W_i\), with \(q_i\) qubits, we construct a set
\(
D_i \subseteq V_D,
|D_i| = q_i,
\)
such that the collection \(D_1,\dots,D_K\) is pairwise disjoint and each \(D_i\) remains spatially localized in \(G\). We design Algorithm~\ref{alg:greedy-compact-partition} to solve this problem. Each workload is first seeded at the currently unoccupied data vertex closest to its fixed magic-state port \(m_i\). The algorithm then grows the clusters in round-robin order: as long as some workload has not yet received all \(q_i\) data vertices, the inner scan over \(i=1,\dots,K\) assigns at most one additional vertex to each unfinished cluster. For any workload with \(|D_i|<q_i\), the next assigned vertex is the currently unoccupied data vertex minimizing its graph distance to the existing cluster \(D_i\). 
We precompute the graph distances for the given architecture abstraction, therefore making the runtime complexity of Algorithm \ref{alg:greedy-compact-partition} $O(Q|V_D|)$ with $Q = \sum_{i=1}^K q_i$.

\begin{figure}
    \centering
    \includegraphics[width=1\linewidth]{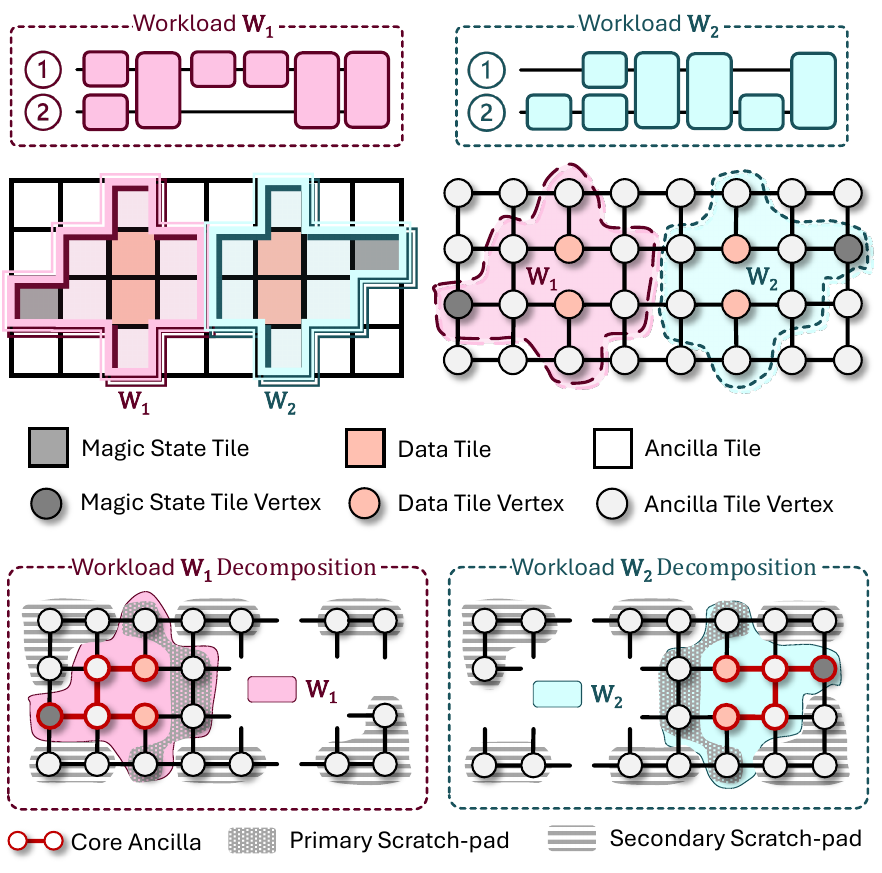}
    \caption{\textbf{Workload Placement \& Abstraction.} A diagrammatic representation of placing two workloads $W_1$ and $W_2$ on the floorplan. In the workload decomposition, we show the core ancilla region of a workload, the primary scratchpad, and the secondary scratchpad allocated to a workload and shared by the other.}
    \label{fig:3b_mapping_workloads}
\end{figure}
\noindent\textbf{Constructing the workload subgraph: }
Once the compact data partition \(D_1,\dots,D_K\) has been fixed, we construct a connected resident subgraph joining the selected data vertices \(D_i\) to the fixed magic-state port \(m_i\) using ancilla vertices. For workload \(W_i\), this amounts to finding
\(
A_i \subseteq V_A
\)
of minimum cardinality such that the induced subgraph on
\(
D_i \cup A_i \cup \{m_i\}
\)
is connected. This is a well-researched problem in lattice surgery scheduling \cite{silva}\cite{dascot}\cite{hofmeyr2026schedulinglatticesurgerymagic}, and we use a Steiner-style heuristic to frame the subgraph with the terminals as the data vertices and the fixed port, and ancilla vertices as Steiner nodes.
We construct the core incrementally. Starting from the singleton root \(\{m_i\}\), Algorithm \ref{alg:incremental-rooted-core} repeatedly attaches one not-yet-connected data vertex of \(D_i\) to the current core using a path that introduces the fewest new ancilla vertices. 
This stage can be implemented in
\(
O\!\bigl(Q(|V|+|E|)\bigr)
\)
time via repeated multi-source \(0\)-\(1\) shortest-path computations. The terminals of the Steiner tree are
\(
R_i = D_i \cup \{m_i\},
\)
and each available ancilla vertex has a unit cost. By applying the standard 
vertex-splitting transformation, this node-weighted instance can be reduced 
to an edge-weighted Steiner-tree instance. Algorithm \ref{alg:incremental-rooted-core} then corresponds to 
the shortest-path Steiner heuristic, which grows a tree by repeatedly 
connecting the closest unconnected terminal to the current partial tree. Thus, for a fixed single-workload connection subproblem, Algorithm \ref{alg:incremental-rooted-core} 
inherits the standard $(2-2/|R_i|)$ approximation guarantee of the 
shortest-path Steiner heuristic \cite{takahashi1980approximate}.
After the compact data partition and rooted core construction have been completed, the resulting family
\(
\mathcal{H}=\{H_1,\dots,H_K\}
\)
must satisfy the pairwise vertex-disjointness condition
\(
V_i \cap V_j = \varnothing,
\forall\, i\neq j.
\)

\begin{algorithm}[t]
\caption{Incremental Rooted Core Construction}
\label{alg:incremental-rooted-core}
\fontsize{8.5pt}{9.5pt}\selectfont
\begin{algorithmic}[1]
\Require Floorplan graph \(G=(V,E)\) with ancilla vertices \(V_A\); data clusters \(D_1,\dots,D_K\); fixed ports \(m_1,\dots,m_K\)
\Ensure Connected core subgraphs \(H_i=(V_i,E_i)\), primary scratchpads \(P_i\), and secondary scratchpad \(S\), if successful
\State \(U_A \gets V_A\)
\For{\(i=1\) to \(K\)}
    \State \(C_i \gets \{m_i\}\), \quad \(E_i \gets \varnothing\), \quad \(A_i \gets \varnothing\), \quad \(R_i \gets D_i\)
    \While{\(R_i \neq \varnothing\)}
        \If{no path in \(G[C_i \cup U_A \cup D_i]\) connects any \(d\in R_i\) to \(C_i\)}
            \State \Return Failure
        \EndIf
        \State choose \((d_i^\star,P_i^\star)\) minimizing \(|V(P_i^\star)\cap U_A|\),
        \Statex \hspace{\algorithmicindent} where \(d_i^\star\in R_i\) and \(P_i^\star\) is a path in \(G[C_i \cup U_A \cup D_i]\) from \(d_i^\star\) to \(C_i\)
        \State \(C_i \gets C_i \cup V(P_i^\star)\)
        \State \(E_i \gets E_i \cup E(P_i^\star)\)
        \State \(A_i \gets A_i \cup (V(P_i^\star)\cap V_A)\)
        \State \(U_A \gets U_A \setminus (V(P_i^\star)\cap V_A)\)
        \State \(R_i \gets D_i \setminus C_i\)
    \EndWhile
    \State \(V_i \gets D_i \cup A_i \cup \{m_i\}\), \quad \(H_i \gets (V_i,E_i)\)
\EndFor
\State \(S \gets U_A\)
\For{\(i=1\) to \(K\)}
    \State \(P_i \gets \{a\in S : d_G(a,V_i)=1\}\)
\EndFor
\State \Return \((H_1,\dots,H_K), (P_1,\dots,P_K), S\)
\end{algorithmic}
\end{algorithm}



We observe this scenario in Fig. \ref{fig:3b_mapping_workloads}. For example, workload $W_1$ is allocated the pink-shaded region of the floorplan by choosing the data vertices first, constructing the subgraph (Fig. \ref{fig:3b_mapping_workloads}), allocating the primary scratchpad, and then defining the secondary scratchpad.
Taken together, Algorithms~\ref{alg:greedy-compact-partition} and~\ref{alg:incremental-rooted-core} define the baseline static allocator used throughout the remainder of the paper. We incorporate targeted modifications to cater to different resource-limited scenarios in the following sections of the paper.

\subsection{Resource-Limited Scenarios}\label{subsec:limitedresource}

\begin{figure*}
    \centering
    \includegraphics[width=1\linewidth]{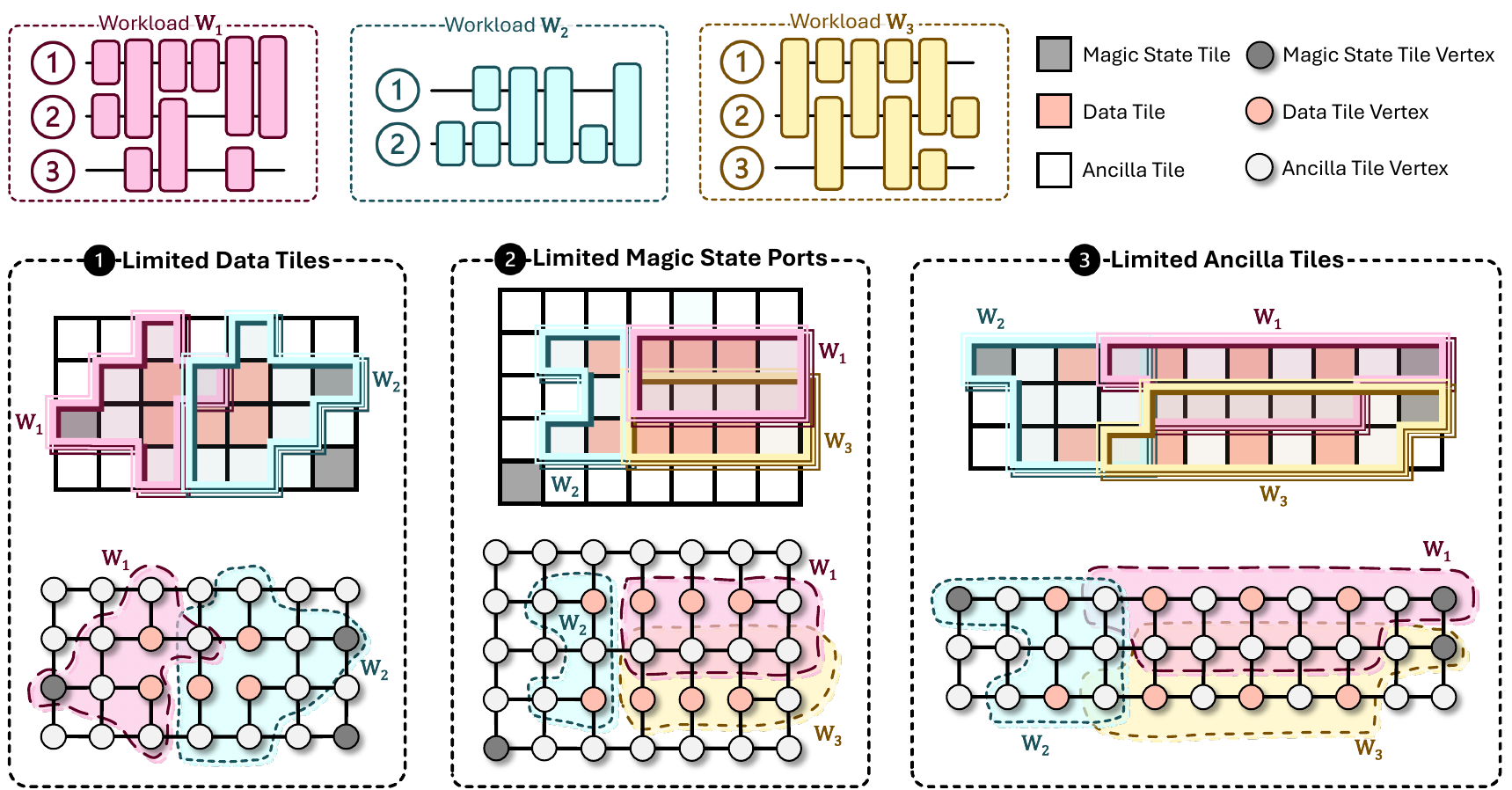}
    \caption{\textbf{Resource Arbitration.} In this diagram, we represent an example of the three different resource-limited scenarios that can appear during concurrent execution of independent workloads. (1) depicts the scenario where the data tiles are limited. We see that $W_1$ and $W_2$ are scheduled, whereas $W_3$ cannot be scheduled due to a lack of available data tiles. (2) represents the case of having limited magic state ports. All the workloads can be executed simultaneously, but they share a single magic state port that is arbitrated based on the latency to deliver the magic state to the workload. (3) shows the scenario where the architecture is data-tile dense. We observe that the ancilla regions (all of the core, primary, and secondary ancilla) overlap and are shared based on a runtime scheduler.  }
    \label{fig:3c_resource_limited}
\end{figure*}

In this section, we consider scenarios in which one or more architecture resources are no longer sufficient to support the full static allocation.

\noindent\textbf{Scenario 1: Limited Data Tiles:}
We first consider the regime in which data tiles are the limiting resource, while ancilla tiles and magic-state ports remain sufficient for any workload that is admitted (Fig. \ref{fig:3c_resource_limited}(a)). The placement problem, therefore, shifts from constructing a simultaneous allocation for all workloads to selecting a subset of waiting workloads whose data demands can be embedded in the currently free floorplan without excessively \emph{fragmenting} the residual data-tile space. 
Let
\(
U_D^{\mathrm{free}}(t)\subseteq V_D
\)
denote the set of data vertices that are unoccupied at scheduling time \(t\). A waiting workload \(W_i\), with demand \(q_i\) and designated port \(m_i\), is data-feasible at time \(t\) if there exists a compact cluster
\(
D_i \subseteq U_D^{\mathrm{free}}(t),
|D_i|=q_i,
\)
that can be generated by Algorithm~\ref{alg:greedy-compact-partition}, now restricted to the currently free data set. 
Therefore, the admissible data vertices is narrowed from \(V_D\) to \(U_D^{\mathrm{free}}(t)\).

The essential modification in the data-limited scenario lies in the order in which this compact-cluster construction is applied. 
Under data scarcity, we admit only workloads for which a compact free cluster of size \(q_i\) is currently available. Feasible workloads are prioritized by decreasing size, with ties resolved using a best-fit rule that minimizes residual slack in the host free region.
Formally, for each currently feasible workload \(W_i\), let \(\sigma_i(t)\) denote the number of unused data vertices left in the smallest free region that can host a compact cluster of size \(q_i\). The admission order is determined lexicographically by
\(
(-q_i,\sigma_i(t)).
\)
After selecting a workload \(W_i^\star\), its assigned cluster \(D_i^\star\) is reserved, the free set is updated by
\(
U_D^{\mathrm{free}}(t)\gets U_D^{\mathrm{free}}(t)\setminus D_i^\star,
\)
and feasibility is recomputed for the remaining waiting workloads. The output of this step is therefore a maximal admitted subset together with pairwise disjoint compact data clusters for the admitted jobs.



\noindent\textbf{Scenario 2: Limited Magic State Ports:}
We next consider the regime where the number of magic-state ports is limited. In this setting, the fixed port \(m_i\) is no longer treated as a permanently reserved vertex of the resident workload subgraph (Fig. \ref{fig:3c_resource_limited}(b)). Instead, the resident allocation of workload \(W_i\) is defined by
\(
H_i=(V_i,E_i), V_i=D_i\cup A_i,
\)
where \(D_i\subseteq V_D\) is the compact data cluster. Magic-state ports are instead modeled as shared external service resources that are accessed on demand during execution. 
Under this abstraction, Algorithm~\ref{alg:incremental-rooted-core} is modified by removing the fixed port from the connectivity objective. Therefore, the initial seed is set to some node $r_i$, $r_i\in D_i$ instead of $m_i$, and the set of remaining terminals is initialized as \(R_i\gets D_i\setminus\{r_i\}\), with the rest of the algorithm and definitions of primary and secondary scratchpad remaining unchanged. The final resident vertex set is therefore \(V_i=D_i\cup A_i\).

Magic-state delivery is handled dynamically. For each port \(p\in V_M\), let \(T_{\mathrm{init}}(p)\) denote its initial warmup latency to produce magic states and \(T_{\mathrm{prep}}(p)\) its production latency. For each \(W_i\), we define \(L(i,p)\) as the shortest distance from the port \(p\) to \(H_i\) (since moving one tile takes one surface code cycle). Each candidate port is evaluated using
\(
T_{\mathrm{deliver}}(i,p)
=
T_{\mathrm{init}}(p)+T_{\mathrm{prep}}(p)+L(i,p),
\)
and the request is assigned to a port that minimizes $T_{\mathrm{deliver}}(i,p)$. If multiple workloads contend for the same next service slot of a port \(p\), arbitration is performed using the lexicographic priority
\(
(-a_i(t),\,L(i,p),\,i),
\)
where \(a_i(t)=t-t_i^{\mathrm{req}}\) is the waiting age of the request issued by workload \(W_i\). Thus, older requests are served first, and ties are broken by lower delivery latency first.

\begin{figure}
    \centering
    \includegraphics[width=0.9\linewidth]{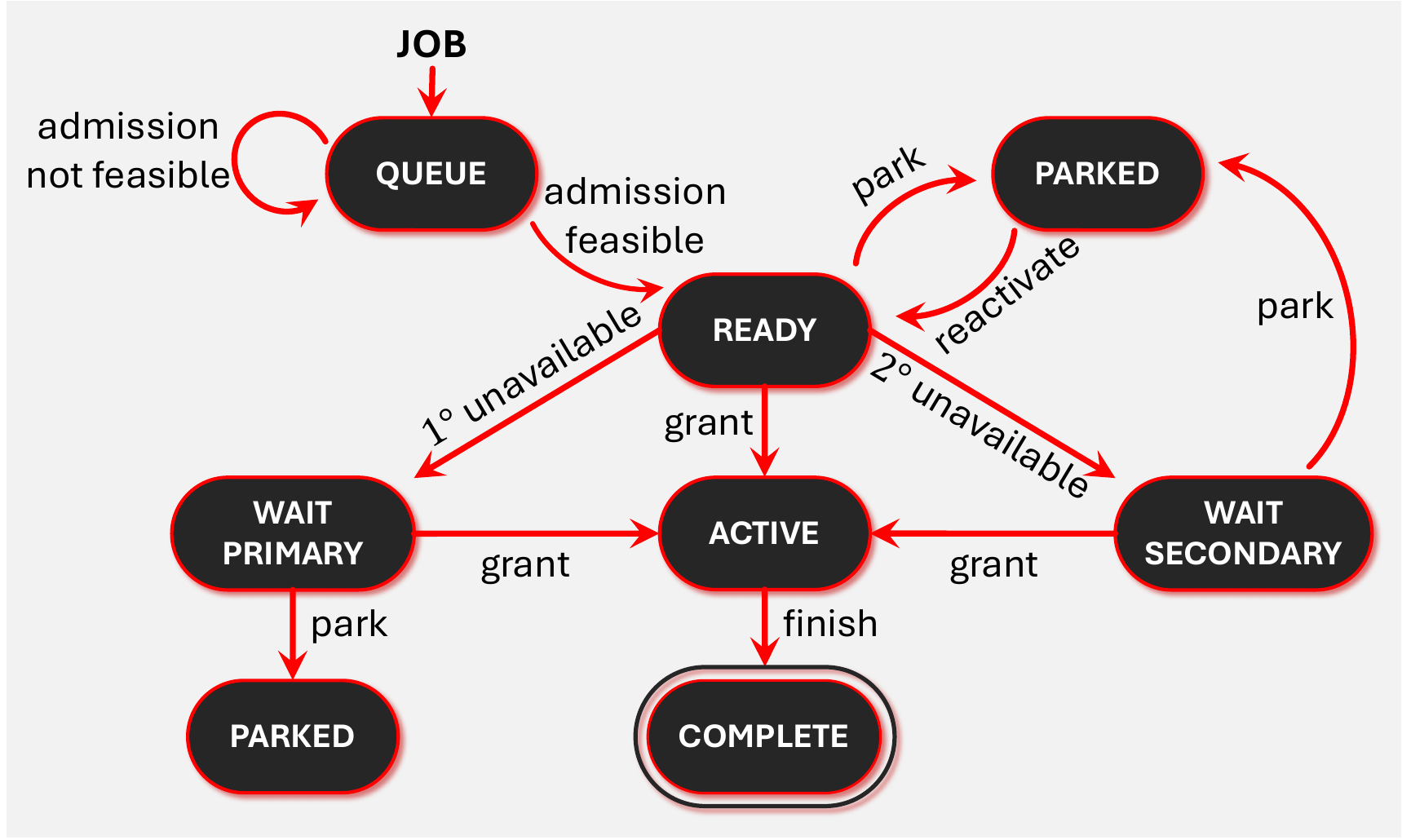}
    \caption{\textbf{The state machine governing the resource arbitration for limited resources under online job scheduling.} The start state is the \textsc{Queue} where the job enters, and the end state is the \textsc{Complete}. A job is held at the \textsc{Parked} state when the ancilla resources are unavailable, and enters the \textsc{Ready} state to get executed.}
    \label{fig:state_machines_multiprog}
\end{figure}

\noindent\textbf{Scenario 3: Limited Ancilla Tiles:}
We finally consider the scenario in which ancilla tiles are the limiting resource (Fig. \ref{fig:3c_resource_limited}(c)). We tackle this problem by introducing a hierarchical sharing policy: secondary scratchpad ancilla is the most elastic resource, primary scratchpad ancilla is spatially local and conflict-prone, and core ancilla is the least elastic and is reclaimed only under more severe scarcity. 
For each workload \(W_i\), let
\(
A_i^{\min}\subseteq V_A
\)
denote the minimum core ancilla. This is the only ancilla allocation treated as structurally pinned while the workload remains resident. All ancilla demand beyond \(A_i^{\min}\) is phase-specific and elastic. At each scheduler boundary, the runtime exposes only the next executable phase of each resident workload through a descriptor
\(
\phi_i=(\tau_i,\,R_i^{\mathrm{prim}},\,\delta_i^{\mathrm{sec}},\,\ell_i,\,\beta_i),
\)
where \(\tau_i\) is the next executable phase, \(R_i^{\mathrm{prim}}\subseteq V_A\) is the primary-scratchpad region, \(\delta_i^{\mathrm{sec}}\ge 0\) is the required amount of secondary scratchpad ancilla, \(\ell_i\) is the estimated phase duration, and \(\beta_i\in\{0,1\}\) indicates whether the phase is blocking. Ancilla is granted only at phase boundaries, since workload execution is \emph{non-preemptive}. Once \(\phi_i\) is granted, \(R_i^{\mathrm{prim}}\) and \(\delta_i^{\mathrm{sec}}\) remain committed to \(W_i\) until that phase completes. A working state machine of the scheduler is shown in Fig. \ref{fig:state_machines_multiprog}. The scheduler therefore has only three control actions: it may grant the next phase of a workload if the required ancilla is currently available, delay the phase if the required ancilla is unavailable or conflicts with a higher-priority grant, or park the workload at a safe boundary by releasing its core ancilla while keeping its data placement fixed.
At each scheduler boundary, a feasible resident set under the available core-ancilla budget is determined. A set of workloads may remain resident simultaneously only if
\(
\sum_{W_i\in\mathcal{R}} |A_i^{\min}| \le C_A,\) for $C_A$ being the core ancilla for the workload $W_i$. A workload is \textsc{Parked} by releasing their core ancilla (while the data tiles remain pinned) if this condition fails. Furthermore, it is delayed till this condition is met.
If the free secondary pool is too small to satisfy all current \(\delta_i^{\mathrm{sec}}\), then affected workloads wait in \textsc{WaitSecondary}, and if they remain feasible but some requested regions \(R_i^{\mathrm{prim}}\) overlap, the blocked workloads remain in \textsc{WaitPrimary} until the required local region becomes available.  



\subsection{Online Arrival of Workloads}

We now extend the scheduler to cases where workloads arrive over time rather than being given as a fixed batch (online arrival). We consider all the limited-resource scenarios together for this exploration. Therefore, the online job scheduler inherits the same structure as the limited-ancilla scheduler Fig. \ref{fig:state_machines_multiprog}. The control of this scheduler is resolved in the order of mapping workload to data tiles, resident-ancilla set control, and runtime arbitration. 
For each waiting workload $W_i$ in the job queue, the scheduler first tests spatial feasibility by invoking the same compact-cluster growth primitive used in Algorithm \ref{alg:greedy-compact-partition}.
If this call fails, $W_i$ is not admissible at time $t$. If a feasible compact cluster exists, the scheduler moves the job to the \textsc{Ready} state, and applies a residency-safety check to ensure the commitment of the workload if and only if the core ancilla is available and the scratchpads are available or can be shared. Owing to the lexicographic ordering $\phi_i$ (discussed earlier), workloads are preferred that do not fragment the free data tile space (qubit count is less than the available contiguous data tile region), larger workloads, and older workloads in \textsc{Parked}, in this order. Once the workload is admitted into the floorplan, the scheduler behaves like the limited ancilla scheduler. It computes the phase of the workload, puts it in the \textsc{Ready} or one of the \textsc{WaitPrimary} or \textsc{WaitSecondary} states based on the core ancilla demand, following up with the arbitration of magic state ports based on $T_{deliver}(i,p)$ of each workload $W_i$, and executing them once it has a set of primary and secondary scratchpads available.

%% file: 4_cultivation.tex
\subsection{Extension to Cultivation-Enabled Architectures}
\label{sec:cultivation-enabled}
\begin{figure}
    \centering
    \includegraphics[width=1\linewidth]{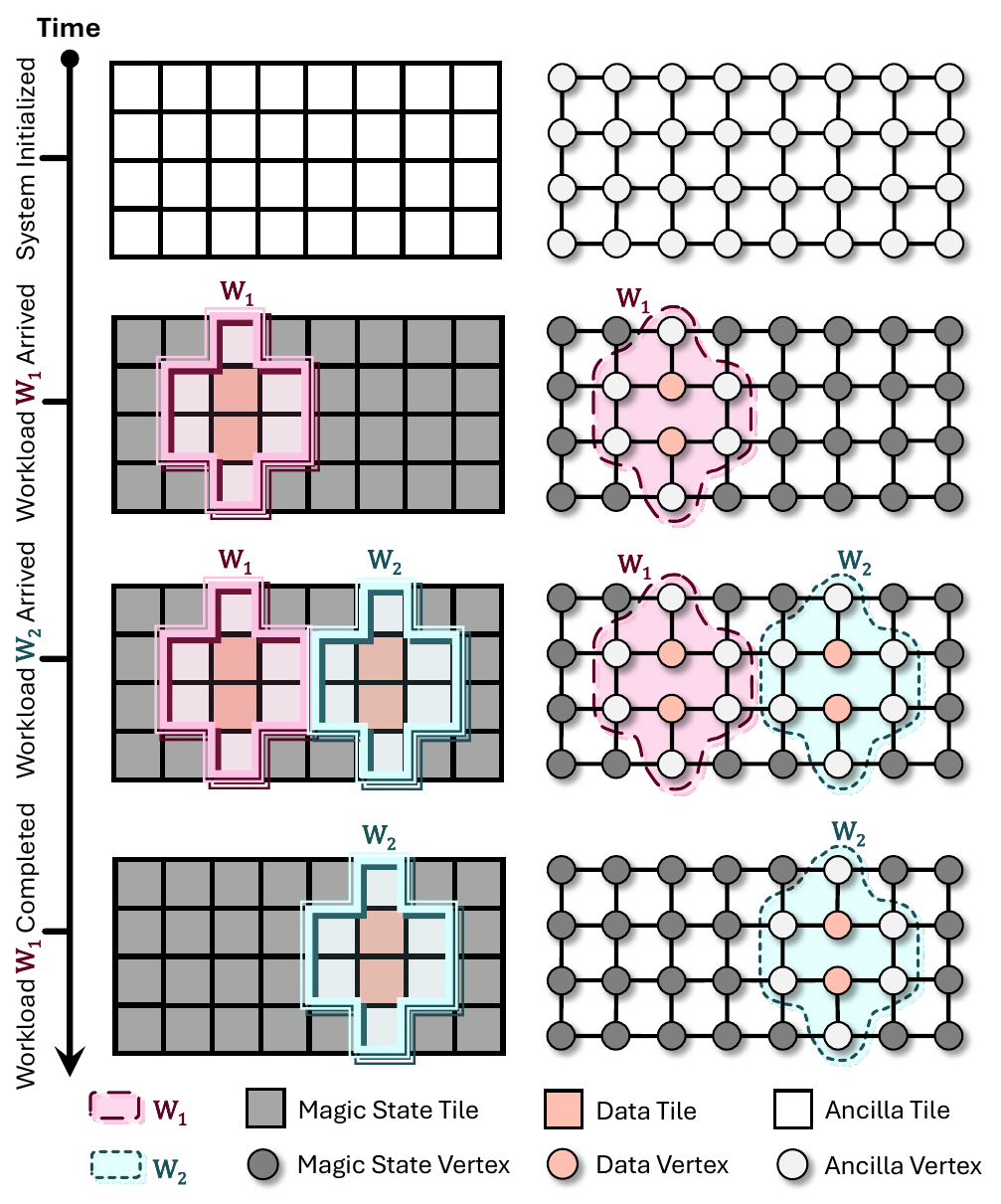}
    \caption{\textbf{Scheduling Cultivation.} A time evolution diagram of the magic state cultivation-enabled architecture. $W_1$ arrives after system initialization, occupies the two data tiles, and the rest of the ancilla (barring the core and primary scratchpad) is cultivating. When $W_2$ arrives, the ancilla demand for measurement and routing increases, and the required ancilla tiles get removed from cultivation. On completion of $W_1$, the previously occupied ancilla tiles are reclaimed and start cultivating.}
    \label{fig:cultivation}

\end{figure}
\begin{figure}
    \centering
    \includegraphics[width=0.9\linewidth]{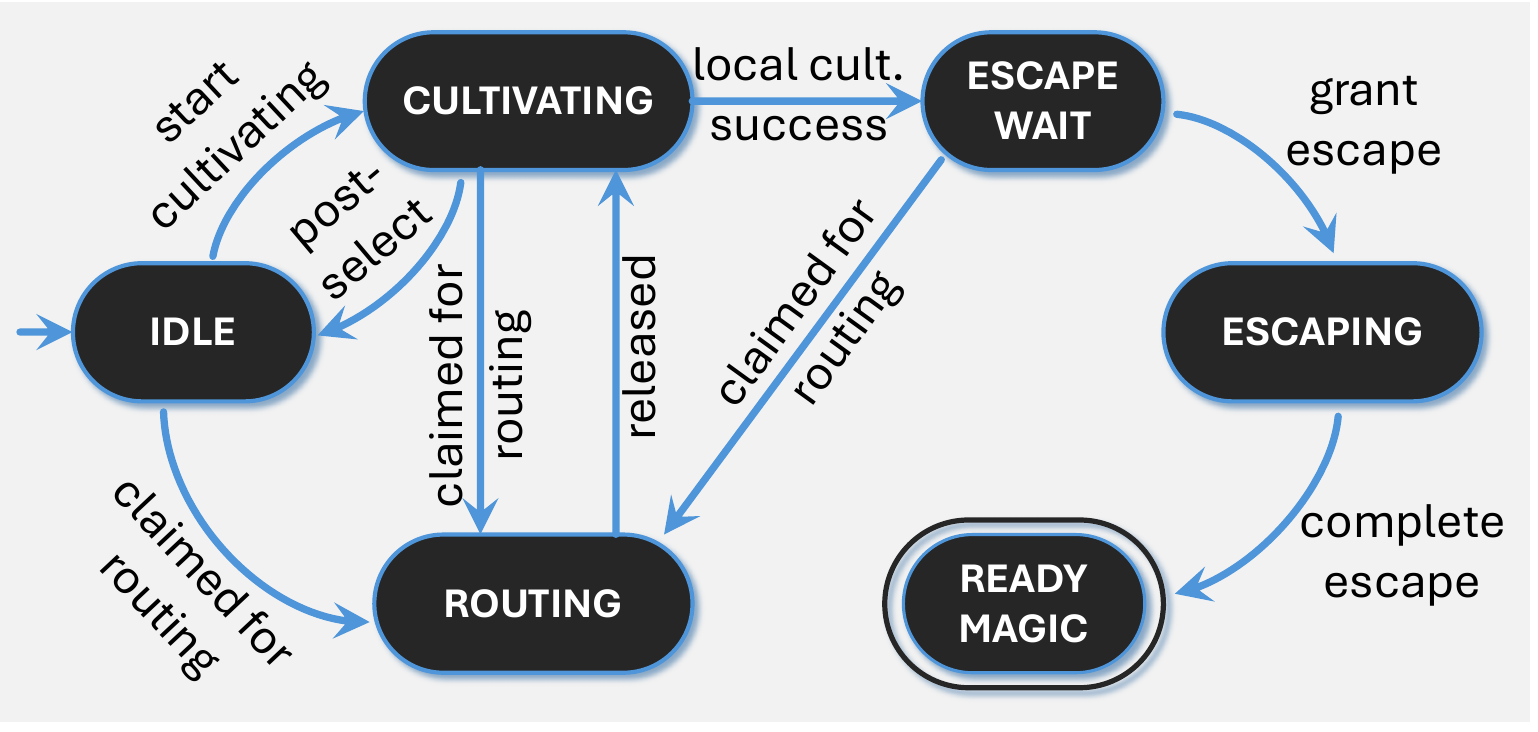}
    \caption{\textbf{The state machine demonstrating the magic state cultivation abstraction in the floorplan.} An ancilla tile, when idle, can either be claimed for \textsc{Routing} or be sent for \textsc{Cultivating}. If the cultivating magic state gets post-selected during cultivation, it gets back to the idle state; otherwise, it escapes when it reaches the desired fault distance and is ready to measure out the $T$ gate. However, a claim for routing is given priority, and whenever a shared, cultivating ancilla tile is needed for routing, it is reclaimed.}
    \label{fig:state_machines_cultivation}
    \vspace{-10pt}
\end{figure}

In this section, we explore the scheduling of multiple workloads using magic state cultivation \cite{hofmeyr2026schedulinglatticesurgerymagic}. Recent work has shown that magic-state cultivation can generate high-fidelity magic states with substantially lower qubit overhead than conventional distillation-based approaches.\cite{gidney2024magicstatecultivationgrowing}\cite{rosenfeld2025magicstatecultivationsuperconducting}. Therefore, we follow up on the magic state cultivation abstraction defined in \cite{hofmeyr2026schedulinglatticesurgerymagic}, and redefine our floorplan abstraction as an undirected, unweighted graph
\(
G = (V,E), V = V_D \sqcup V_A,
\)
where \(V_D\) and \(V_A\) denote the data-tile and ancilla-tile vertices, respectively. There is no distinguished set \(V_M\) of fixed magic-state ports anymore. The resident allocation for a workload $W_i$ becomes $H_i = (V_i,E_i), V_i = D_i \cup A_i$, such that \(D_i \subseteq V_D\), \(A_i \subseteq V_A\), \(|D_i| = q_i\), and \(H_i\) is connected. We consider all ancilla tiles to be in the cultivation stage \cite{gidney2024magicstatecultivationgrowing} when not participating in routing or measurement, with the desired fault distance being the upper bound of the surface code distance for the tile. Consider the \emph{example} in Fig. \ref{fig:cultivation}. We demonstrate the arrival of jobs and the changing state of the floorplan under magic state cultivation. At \emph{System Initialization}, the $n\times n$ grid of qubits is divided into tiles for surface code patch integration. On the arrival of workload $W_1$, the data tiles are formed greedily (Algorithm \ref{alg:greedy-compact-partition}), and the subgraph is determined (Algorithm \ref{alg:incremental-rooted-core}) with the primary scratchpad in place. Since cultivation can be terminated and restarted at any point in time \cite{gidney2024magicstatecultivationgrowing}\cite{hofmeyr2026schedulinglatticesurgerymagic}, all other tiles start cultivating magic states. This ensures that no ancilla tile is idle. When $W_2$ arrives, the set of tiles required by $W_2$ is computed and terminated from cultivation. During the execution of $W_1$, the ready magic state from the surrounding cultivation patches is used for implementing the $T$ gates, and on completion, the tiles occupied by $W_1$ restart cultivating magic states.



To keep the runtime abstraction compact, for each ancilla tile, we define an arbitration scheduler (Fig. \ref{fig:state_machines_cultivation}) to efficiently cultivate and schedule the magic states when they are ready. The key idea is to use the ancilla tiles for cultivating when they are not used for routing the data patches. 
Magic-state availability is therefore represented by a dynamic ready set rather than by port service times. Let
\(
T(t) := \{a  : \sigma_a(t)=\textsc{ReadyMagic}\}
\)
denote the set of ancilla tiles currently holding usable magic states ($\sigma_a(t)$ is the phase of the tile $a$ at time $t$).The scheduler-visible next-phase descriptor of workload \(W_i\) becomes
\(
\widehat{\phi}_i =
\bigl(\tau_i, R_i^{\mathrm{prim}}, \delta_i^{\mathrm{sec}}, \mu_i, \ell_i, \beta_i \bigr),
\)
where \(\tau_i\) is the next executable phase, \(R_i^{\mathrm{prim}} \subseteq V_A\) is the required local primary-scratchpad region, \(\delta_i^{\mathrm{sec}} \ge 0\) is the required amount of secondary ancilla, \(\mu_i \ge 0\) is the number of ready magic states consumed by the phase, \(\ell_i\) is the estimated phase duration, and \(\beta_i \in \{0,1\}\) is the blocking indicator. When \(\mu_i > 0\), the scheduler must choose a set
\(
M_i(t) \subseteq T(t), |M_i(t)|=\mu_i,
\)
of ready-magic tiles to assign to the phase. Any additional local routing needed to incorporate those assigned tiles into the lattice-surgery operation is accounted for inside \(R_i^{\mathrm{prim}}\) and \(\delta_i^{\mathrm{sec}}\). 

%% file: 5_evaluation.tex
\section{Evaluation}\label{sec:eval}
We conduct detailed numerical simulations to evaluate the efficacy of our proposed multiprogramming framework.
\noindent\textbf{Benchmarking: }We use a custom benchmark suite: A synthetic set of Clifford+$T$ circuits with randomized size and $T$-gate density. This is done due to the lack of FTQC benchmarks that cater to the problem of multiprogramming in large-scale quantum computing \cite{Wakizaka_2025}. Each synthetic workload is parameterized by the number of logical qubits $n$, the number of circuit columns $c$, and the total number of $T$-gates. The $T$-gates are sampled uniformly across the $n \times c$ qubit-column grid, with at most one $T$-gate per qubit in any column. For each selected $T$-gate location, the rotation axis is drawn from $\{X,Y,Z\}$, and the column-level global phase is sampled independently from a Bernoulli distribution with $p=0.5$. Unless stated otherwise, we use the default parameters defined in Table \ref{tab:param}. For the hardware abstraction, we assume the execution on a floorplan with 50\% density of data tiles, which is considered by prior research to be densest for it to be \emph{Immediate-Operation} capable \cite{ueno2024highperformancescalablefaulttolerantquantum}.
We divide the workload mix into 4 categories: (i) \emph{Small} comprising 80\% workloads in the small category, and rest random; (ii) \emph{Medium} comprising 80\% workloads in the medium category; (iii) \emph{Big} comprising 80\% workloads in the big category; and (iv) \emph{Balanced} comprising an equal distribution of the workloads. 
\begin{table}[ht]\label{tab:param}
\centering
\caption{Experimental Parameters}
\fontsize{7.5pt}{8.5pt}\selectfont
\begin{tabular}{cc}
\toprule
Parameter                         & Value                                 \\  
\midrule
Floorplan dimension               & 20x12                                 \\
Data tile density                 & 50\%                                  \\
\# magic state ports              & 50                                    \\
Magic state Distillation protocol & 15-to-1                               \\
Workload count                    & 100                                   \\
T-depth                           & 10-1000                               \\
Circuit type (qubit count)        & Small/Medium/Big (10-20/40-60/60-100) \\
Arrival model                     & Online                                \\
Implementation HW                 & Intel Core i9-13900K CPU (4.8GHz)    \\
\bottomrule
\end{tabular}
\end{table}

We attempt to answer the following research questions through our evaluation:
\begin{itemize}
    \item \textbf{RQ1: }How does our framework perform with respect to prior research and random policies?
    \item \textbf{RQ2: }Which configuration contributes to a net performance gain in our proposed framework?    
    \item \textbf{RQ3: }How is the Quality of Service in the framework under a heavily resource-constrained scenario?
    \item \textbf{RQ4: }How does the framework work in case of cultivation-enabled floorplans?
\end{itemize}

\subsection{(RQ1) Comparative effectiveness}

\begin{figure}
    \centering
    \includegraphics[width=1\linewidth]{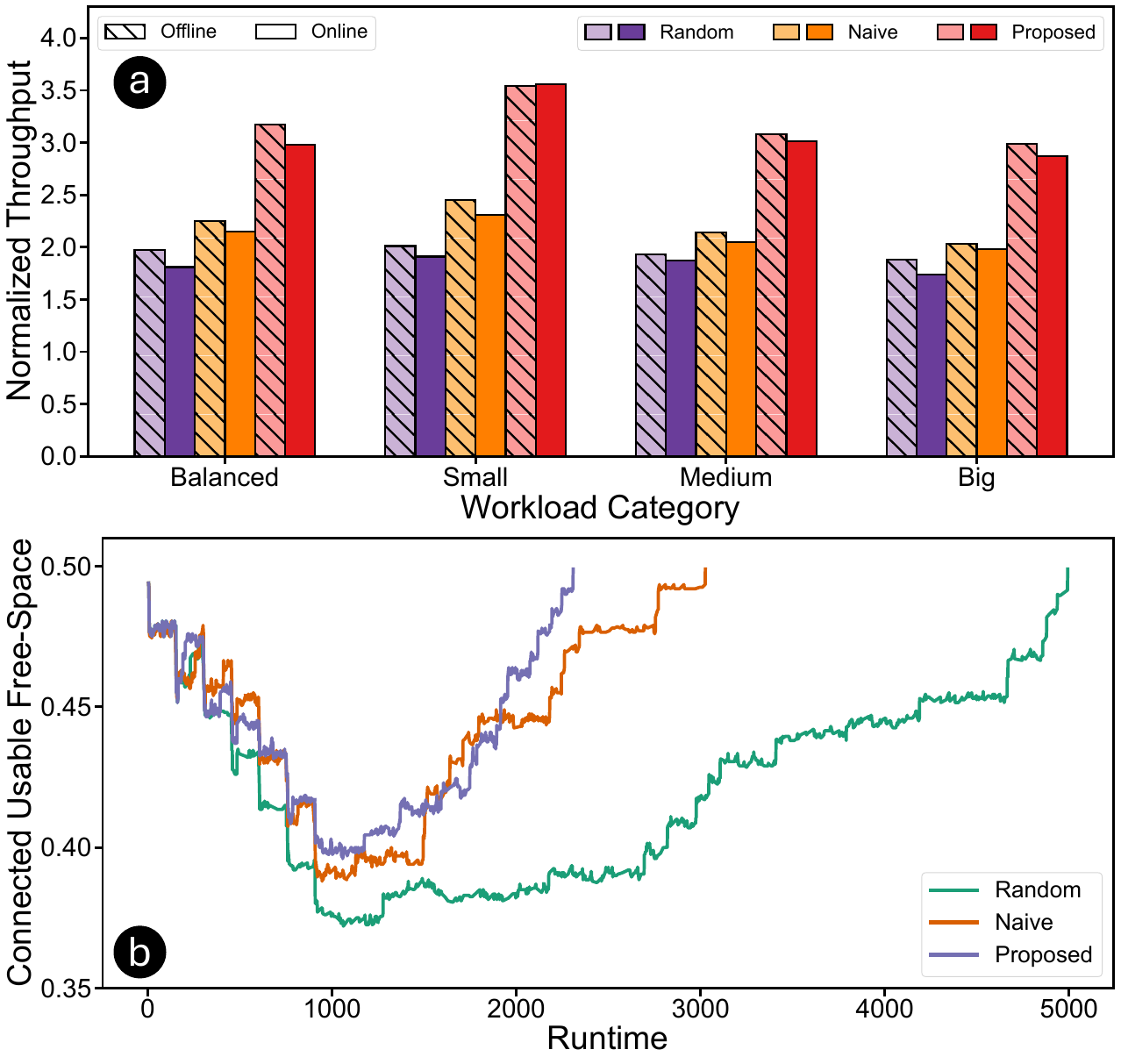}
    \caption{\textbf{Plots describing the RQ1.} In (a) we observe the normalized throughput against random and naive baselines. In (b), we show the system trace of running a \emph{Balanced} mix of workloads, depicting the total usable free space in the floorplan. Time is abstracted as 1 time step = 1 surface code cycle.}
    \label{fig:rq1_plots}
\end{figure}

We first compare the proposed framework against the only prior FTQC multiprogramming approach \cite{Wakizaka_2025}. We compare against their Integer Linear Programming-based on cuboid representations (ILP-C) and the Corner Greedy (CG) scheduler, both with defragmentation as their optimization knob enabled. To make an apples-to-apples comparison, we use the default experiment parameters with 300 workloads, and a 25\% data-dense floorplan and compute the speedup and report the results in Table \ref{tab:comp}. We find that our proposed framework provides a $\sim 29\%$ increase in speedup compared to their Corner Greedy implementation and $\sim 34\%$ increase compared to their ILP formulation (both implemented using the defragmentation strategy). To show the effectiveness of our proposed framework, we define a normalized throughput metric 
\[
\eta(T)=\frac{\sum_{W_i} T_i^{\mathrm{solo}}}{T},
\]
where \(T_i^{\mathrm{solo}}\) is the completion time of workload \(W_i\) when executed alone on the same architecture. Further, we evaluate our proposed scheduler against two simpler internal baselines, \emph{naive} and \emph{random}, under identical architectural and workload assumptions. The \emph{naive} baseline retains the same admission policies as the proposed method, but neglects the scheduler ones. The \emph{random} baseline randomly shares resources by selecting among feasible placements without structured optimization. We observe the results in Fig. \ref{fig:rq1_plots}(a). In an offline arrival scenario, we find the highest normalized throughput for the proposed framework (mean: $3.2$ in offline; $3.1$ in online). When compared to the baselines, we find a $\sim43\%$ improvement in mean normalized throughput over the naive policy and $\sim64\%$ over the random placement. We also investigate the change in \emph{usable free space} in the floorplan with time. We demonstrate the results in Fig. \ref{fig:rq1_plots}(b). Let $O_D(t) \subseteq V_D$ and $O_A(t) \subseteq V_A$ denote the data and
ancilla tiles occupied at time $t$, respectively. The set of free vertices is
\(
V^{\mathrm{free}}(t)
=
\bigl(V_D \setminus O_D(t)\bigr)
\cup
\bigl(V_A \setminus O_A(t)\bigr).
\)
Let
\(
G^{\mathrm{free}}(t) = G\bigl[V^{\mathrm{free}}(t)\bigr]
\)
be the subgraph induced by the free vertices, and let
\(
\mathcal{C}(t)=\{C_1(t),C_2(t),\ldots,C_{r(t)}(t)\}
\)
denote the connected components of $G^{\mathrm{free}}(t)$.
We define the largest usable free-space fraction as
\[
C_{\max}(t)
=
\frac{
\max\limits_{C_j(t)\in \mathcal{C}(t)} |C_j(t)|
}{
|V^{\mathrm{free}}(t)|
}.
\]
A larger value of $C_{\max}(t)$ indicates that the remaining free space is concentrated in a large connected region, making it more useful for admitting new workloads. Conversely, a smaller value indicates that the free space is
fragmented across many disconnected components. From Fig. \ref{fig:rq1_plots}(b), we observe that our proposed framework (a) completes the execution earlier than other baselines, and (b) has the highest percentage of \emph{usable free space}, which indicates the efficiency of our approach. 

\begin{rqsummarybox}
    
\textbf{RQ1 Summary: }The proposed scheduler improves system-level normalized throughput over prior FTQC multiprogramming by $\sim 29\%$ and is better than a random placement by preserving compactness and maintaining up to $\sim 10\%$ more contiguous free space during peak execution.
\end{rqsummarybox}

\begin{figure}
    \centering
    \includegraphics[width=1\linewidth]{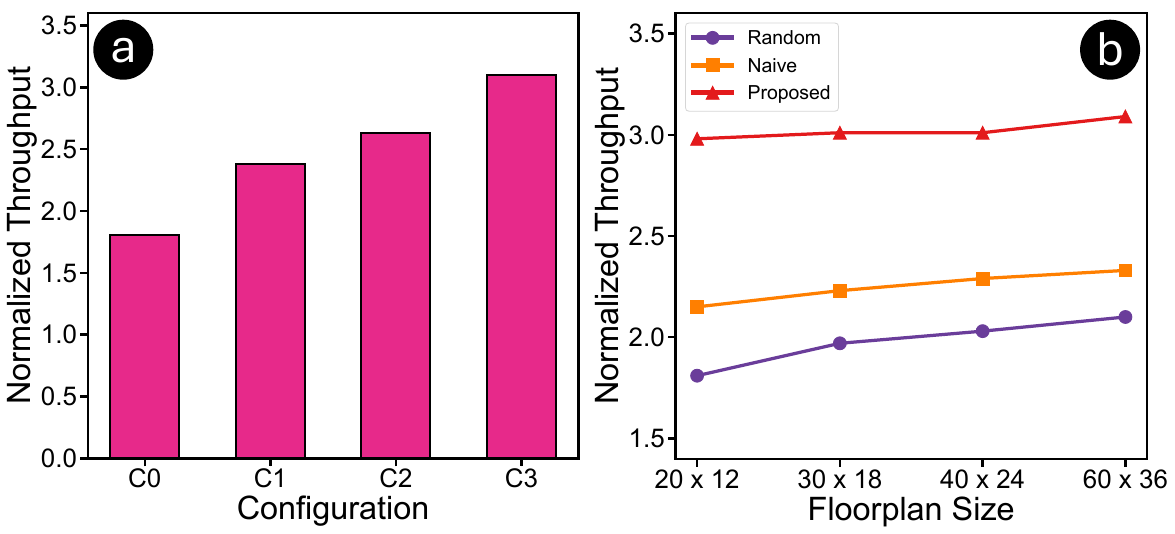}
    \caption{\textbf{Plots describing RQ2.} In (a) we show the ablation study on the four core components of our framework. In (b), we demonstrate the scalability of our framework with an increase in floorplan size.}
    \label{fig:new_rq3_plots}
\end{figure}

\begin{table}[t]
\centering
\caption{Comparison of the proposed idea with prior research}
\begin{tabular}{lccc}
\toprule
Methodology: & ILP-C~\cite{Wakizaka_2025} & CG~\cite{Wakizaka_2025} & Proposed \\
\midrule
Normalized Throughput: & 2.32 & 2.4 & \textbf{3.1} \\
\bottomrule
\end{tabular}
\label{tab:comp}
\end{table}

\subsection{(RQ2) Ablation Study}
To study the contribution of four core components of our framework to performance gains, we design an ablation study with four configurations, $C_0, C_1, C_2, C_3$ (described in Table \ref{tab:config}), built incrementally. We perform the experiments on a \emph{Balanced} workload mix and report the normalized throughput in Fig. \ref{fig:new_rq3_plots}(a). We observe the highest gain while incrementing from $C_0$ to $C_1$ at $\sim31\%$ increase in the normalized throughput, determining that the heuristic algorithms for placement of workloads efficiently contribute most to the improvement in performance, followed by the implementation of online admission ($C_2$ to $C_3$), which contributes to a $\sim17\%$ increase. The hierarchy-aware ancilla arbitration policy ($C_1$ to $C_2$) also contributes to a further $\sim10\%$ gain by reducing ancilla fragmentation. Further, we explore the scalability of our proposed framework to larger floorplans. We increase the size of the floorplan from the default ($20\times12$) to $30\times18$, $40\times24$, and $60\times36$, and run a mix of \emph{Balanced} workloads (of sizes 100, 250, 400, 900 respectively) for the proposed as well as the Random and Naive baselines. We increase the size of the workload mix proportionally to make the comparison non-trivial. As floorplan size grows, reduced ancilla contention lifts all policies; the proposed framework maintains a consistent absolute throughput advantage of $\sim1\times$ over the random baseline across all scales, demonstrating that the framework is scalable with an increase in floorplan size.

\begin{table}[ht]
\centering
\caption{Configurations for Ablation Study}
\begin{tabular}{cc}
\toprule
Configuration & Active Components                     \\
\midrule
$C_0$            & Random placement, FIFO admission      \\
$C_1$           & + Placement policy (Alg. \ref{alg:greedy-compact-partition}, Alg. \ref{alg:incremental-rooted-core})   \\
$C_2$            & + Hierarchy-aware ancilla arbitration \\
$C_3$            & + Online Admission             \\      
\bottomrule
\end{tabular}
\label{tab:config}
\end{table}

\begin{rqsummarybox}
    
\textbf{RQ2 Summary: }The ablation study shows that workload-placement heuristics provide the largest performance contribution, improving normalized throughput by $\sim31\%$, while online admission adds a further $\sim17\%$. We further show that the framework maintains a consistent normalized throughput advantage of $\sim1
\times$ over the random baseline.
\end{rqsummarybox}

\subsection{(RQ3) Quality of Service}
\begin{figure*}
    \centering
    \includegraphics[width=1\linewidth]{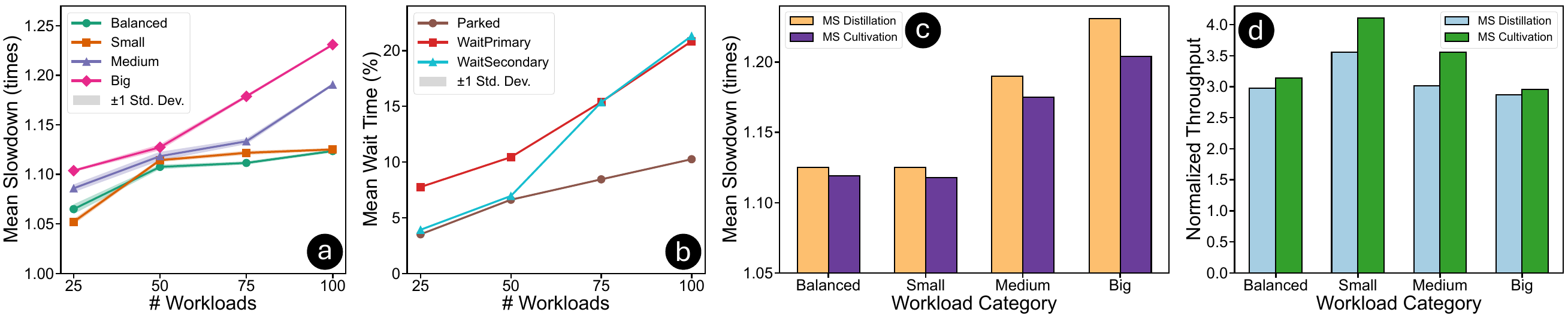}
    \caption{\textbf{Plots describing RQ3 and RQ4.} In (a) we show the mean slowdown in our proposed method with the increase in the number of workloads. In (b) we show the mean wait time as a percentage of the total execution time of the workload with the increase in the number of workloads. In (c) and (d) we demonstrate the reduction in slowdown and increase in normalized throughput, respectively, when using cultivation-enabled architectures.}
    \label{fig:rq2_rq3_plots}
\end{figure*}

We next study the slowdown in individual workload execution due to the wait times incurred at the \textsc{Parked}, \textsc{WaitPrimary}, and \textsc{WaitSecondary} states of the scheduler. Let \(T_i^{\mathrm{solo}}\) denote the execution time of workload \(W_i\) when run alone on the architecture, and let \(T_i^{\mathrm{shared}}\) denote its completion time under concurrent execution, measured from its arrival to completion. The slowdown of \(W_i\) is defined as
\(
S_i
=
\frac{T_i^{\mathrm{shared}}}{T_i^{\mathrm{solo}}}.
\)
A value of \(S_i=1\) indicates no degradation relative to standalone execution, while larger values indicate increased delay due to resource sharing, waiting, or parking. We report the mean slowdown across the four workload mixes across 5 seeds in Fig. \ref{fig:rq2_rq3_plots}(a). We find that the average slowdown is $\sim 1.10\times$ across all workload mixes. We also note that the error bands are tight with a standard deviation of order $10^{-3}$, confirming the stability of the scheduler performance. The slowdowns of the \emph{Big} category diverge further ($\sim1.23\times$ at 100 workloads) with increasing size, demonstrating a greater slowdown due to disproportionate resource accumulation in the floorplan. However, the rest of the workload mix categories remain within $1.13\times$.
Further, we investigate the wait times incurred at different stages by a workload in the online scheduler for the \emph{Balanced} workload mix. In Fig. \ref{fig:rq2_rq3_plots}(b), we report the mean of the time spent by a workload in the states as a percentage of the total time taken for the workload to execute across 5 seeds. We observe that as we increase the number of workloads, the percentage of time spent at these states increases in general, and the wait time at the \textsc{WaitSecondary} state increases from $\sim4\%$ while executing 25 workloads to $\sim21\%$ when we increase the number of workloads to 100. Further, we note that at higher workloads, the wait times for \textsc{WaitSecondary} and \textsc{WaitPrimary} converge, indicating that the dominant bottleneck is the secondary scratchpad for larger numbers of workloads. The error bands are tight in this case as well with standard deviation of the order $0.1\%$.

\begin{rqsummarybox}
    
\textbf{RQ3 Summary: }Under heavily resource-constrained online execution, the proposed scheduler keeps average slowdown close to standalone execution ($\sim1.1\times$) while exposing secondary-scratchpad contention as the dominant source of wait time at higher workload pressure.
\end{rqsummarybox}

\subsection{(RQ4) Cultivation-enabled architectures}
Finally, we compare the framework in the presence of magic state cultivation patches instead of magic state ports. We report the results in Fig. \ref{fig:rq2_rq3_plots}. In our experiments, we consider a $d=23$ surface code architecture with a single magic state port attached to a 15-to-1 distillation protocol, which has an initial start time of 11 surface code cycles \cite{Litinski_2019_msd}, and for cultivation patches, it is 26 surface code cycles \cite{gidney2024magicstatecultivationgrowing}. This makes the achievable fault distance the same ($\sim10^{-12}$) \cite{Litinski_2019_msd} for both configurations. Note that in our hardware abstraction (\ref{subsec:hw-abstraction-notation}), the magic state distillation factory circuitry is not represented as a part of the schedulable floorplan graph $G$, and $V_M$ represents only the magic state port through which the distillation circuit feeds the high-fidelity magic state to the workload for implementation of $T$-gates.
We find from the Fig. \ref{fig:rq2_rq3_plots}(c) that the slowdown reduces slightly (average $\sim1\%$), and from Fig. \ref{fig:rq2_rq3_plots}(d) that the normalized throughput increases (average $\sim10\%$). However, this reduces the resource overhead significantly. The improvement in throughput is caused by the elimination of $T_\mathrm{deliver}(i,p)$ (Section \ref{subsec:limitedresource}), as magic states are now readily available at the nearest ancilla tile. The consequent architectural gain is significant. The 15-to-1 magic state distillation protocol takes 164 extra tiles (153 patches and 11 ancilla tiles). For the $d=23$ surface code implementation, that amounts to a difference of 173,348 qubits, which reduces the resource overhead largely with respect to the upcoming FTQC regime. We also observe from the modest normalized throughput improvement that the cost of occasionally reclaiming cultivating ancilla tiles for routing is lower than the cost of reserving a fixed magic state port.

\begin{rqsummarybox}
    
\textbf{RQ4 Summary: }Cultivation-enabled floorplans improve mean normalized throughput by $\sim10\%$ and slightly reduce slowdown $\sim1\%$ by replacing fixed magic-state ports with opportunistic cultivation on otherwise idle ancilla, while avoiding a dedicated distillation-factory region.
\end{rqsummarybox}

%% file: 6_Discussion.tex
\section{Related Works and Discussion}\label{sec:related}
Multiprogramming in NISQ has been widely studied \cite{nisq-qmp, newqmp, Ohkura_2022, ravi2022adaptivejobresourcemanagement, Niu_2023, giortamis2025qosquantumoperating}. These contributions work toward the optimization of latency and maximization of qubit utilization while trying to minimize the crosstalk overhead and maintain execution fidelity. However, the change in abstraction of hardware and research involving the exploration of the implementation of QEC in architecture led to research in FTQC multiprogramming. Authors in \cite{Wakizaka_2025} introduce the idea of scheduling FT workloads and propose a scheduler based on integer linear programming and optimization knobs to defragment the floorplan during workload execution. Our proposed approach improves on their defragmentation policies by using Algorithm \ref{alg:greedy-compact-partition} to actively avoid fragmentation of the floorplan during the admission phase. Moreover, our paper presents a finer-grained approach to multiprogramming by exploring the directions in efficient resource sharing (data tiles, ancilla, and magic states), online arrival of jobs, and advanced FTQC architecture components like magic state cultivation patches.

%% file: 7_conclusion.tex
\section{Conclusion}\label{sec:conc}

In this paper, we present a multiprogramming framework for surface-code FTQC architectures. We propose (i) a formal definition of the hardware abstraction; (ii) heuristics to place workloads on the architecture; (iii) explore the resource-constrained scenarios and cultivation-enabled architectures. Our numerical simulations provide significant gains over prior work in FTQC multiprogramming.